\documentstyle[twoside, epsf]{article}

\input ibvs2.sty

\begin{document}
\IBVShead{6140}{20 April 2015}

\IBVStitletl{PSN J07285387+3349106 in NGC 2388: an extremely}{ rapidly declining luminous supernova}

\IBVSauth{Tsvetkov, D. Yu.; Volkov, I. M.; Pavlyuk, N. N.}

\IBVSinsto{Sternberg Astronomical Institute, M.V. Lomonosov Moscow
State University, Universitetskij pr.13, 119992 Moscow, Russia,
e-mail: tsvetkov@sai.msu.su}

\IBVSabs{We present CCD BVRI photometry for PSN J07285387+3349106 in NGC 2388}
\IBVSabs{collected from 2015 February 17 until March 17. Decaying by}
\IBVSabs{3.6 mag in the R band in the first 20 days post-maximum, this object}
\IBVSabs{is among the fastest supernovae observed to date}

\SIMBADobj{PSN J07285387+3349106}

\begintext

Supernova in NGC 2388 was discovered by 
Lick Observatory Supernova Search (LOSS) on 2015 February 13.338 UT
at an unfiltered magnitude of 16.9. The variable was designated 
PSN J07285387+3349106 when it was posted at the
Central Bureau's TOCP webpage. The new object was located at
RA=7:28:53.87, DEC=+33:49:10.6 (2000), which is $5''$E and $2''$N from the
center of the host galaxy. Spectroscopic observation on 2015 February 18.79 UT was reported by Ochner et al. (2015). The spectrum was
almost featureless, with strong interstellar Na\,I\,D absorption,
which allowed to estimate a reddening value of $E(B-V)\approx1$ mag.
Unresolved emission lines of He\,I (587.6 and 706.5 nm) were detected,
and a black-body temperature of about 15300 K was inferred for the 
transient.
The characteristics were consistent with those of a very young 
core-collapse SN, and a tentative classification as Type Ibn SN 
was suggested for the object.

We performed photometric observations of PSN J07285387+3349106
in the $BVRI$ bands with the 0.7-m reflector in Moscow (M70), equipped
with Apogee AP-7p CCD camera, and the 1.0-m reflector of Simeiz 
Observatory (S100)
with FLI PL09000 CCD. First images were obtained on February 17, and 
the last ones were collected on March 17, 2015.      
The standard image reductions and photometry were made using 
IRAF\footnote{IRAF is distributed by the National Optical 
Astronomy Observatory,
which is operated by AURA under cooperative agreement with the
National Science Foundation}.
Photometric measurements of the SNe were made relative to 
6 nearby local standard stars using PSF-fitting with IRAF DAOPHOT package. 
The $BV$ and $g'r'i'$ magnitudes of these stars were taken from the
AAVSO Photometric All-Sky Survey (Henden et al., 2012; hereafter, APASS).
The $RI$ magnitudes were computed using relations presented by 
Munari\footnote{\tt https://13378703316991552715.googlegroups.com/}. 
Besides, on one photometric night we calibrated $VR$ magnitudes
of the local standards using data from the M70 telescope. The agreement
between the APASS data and our calibration was excellent, confirming
the reliability of the calibration.
Subtraction of the host galaxy background was necessary for reliable 
photometry, because the SN occurred in one of the spiral arms of the galaxy and 
close to the nucleus. As the SN was practically undetectable in the images 
obtained on March 17, we used these frames to subtract the galaxy 
background from the images collected in February, when the SN was bright.
For the images obtained in March we used the 
SDSS\footnote{\tt http://www.sdss.org} images for 
subtraction. We plan to obtain new deep images of the host galaxy with 
our telescopes and improve the quality of subtraction.
The instrumental magnitudes were transformed to the standard 
Johnson-Cousins system using linear colour equations.
The photometry for the SN is presented in Table 1, the light 
curves are shown in Fig. 1.

\IBVSfig{10cm}{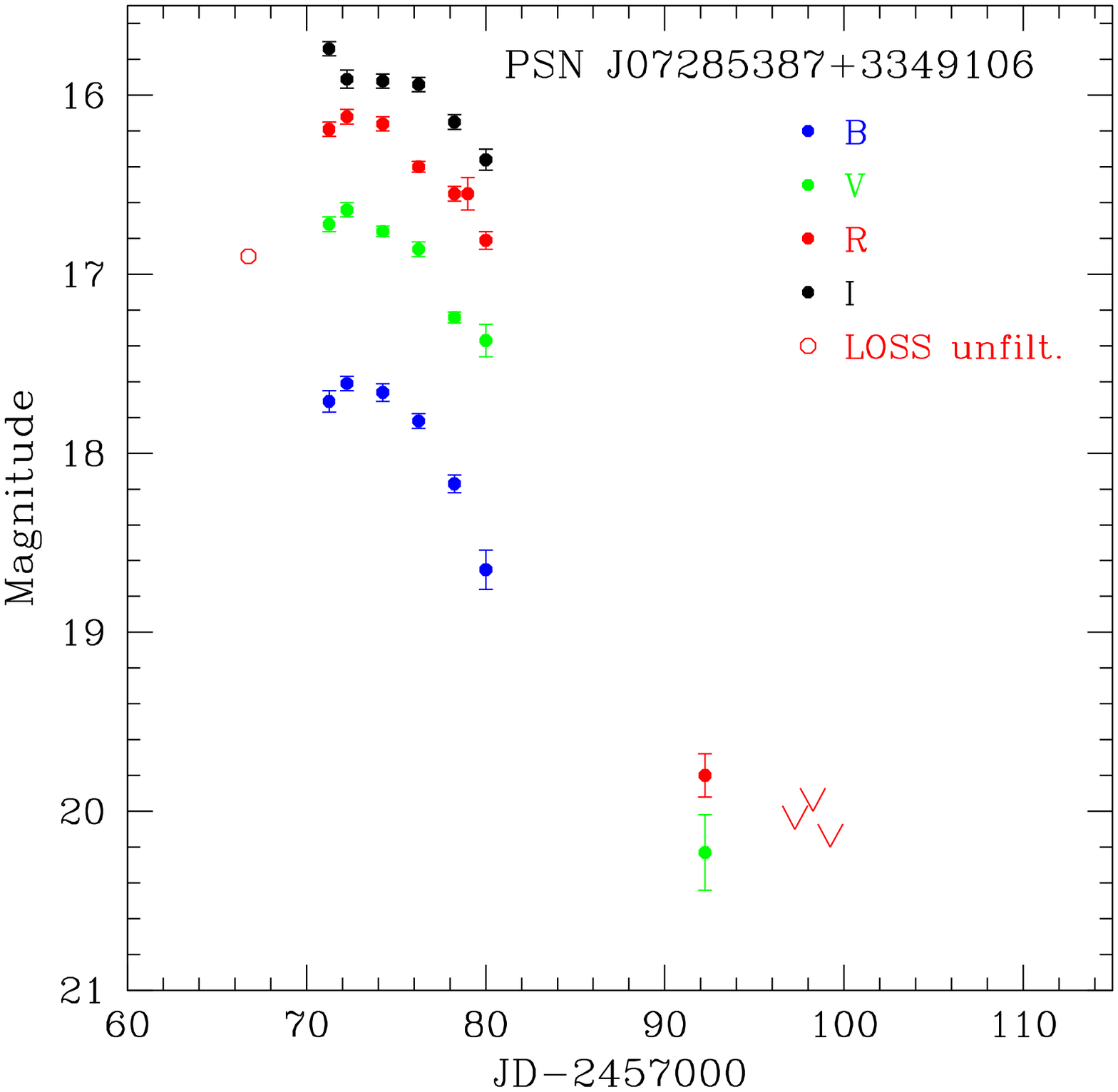}{The $BVRI$ light curves for 
PSN J07285387+3349106.}
\IBVSfigKey{6140-f1.eps}{PSN J07285387+3349106}{light curve}

\begin{table}
\centering
\caption{$BVRI$ photometry of PSN J07285387+3349106}
\begin{tabular}{cccccccccc}
\hline
JD $-$ & $B$ & $\sigma_B$ & $V$ & $\sigma_V$ & $R$ & $\sigma_R$ 
& $I$ & $\sigma_I$ & Telescope \\
2457000 & & & & & & & & & \\ 
\hline
71.22 & 17.71& 0.06&  16.72& 0.04&  16.19 & 0.04 & 15.74& 0.04& M70  \\
72.31 & 17.61& 0.04&  16.64& 0.04&  16.12 & 0.04 & 15.91& 0.05& S100 \\ 
74.31 & 17.66& 0.05&  16.76& 0.03&  16.16 & 0.04 & 15.92& 0.04& S100 \\
76.39 & 17.82& 0.04&  16.86& 0.04&  16.40 & 0.03 & 15.94& 0.04& S100 \\
78.28 & 18.17& 0.05&  17.24& 0.03&  16.55 & 0.04 & 16.15& 0.04& S100 \\
79.15 &      &     &       &     &  16.55 & 0.09 &      &     & M70  \\
80.16 & 18.65& 0.11&  17.37& 0.09&  16.81 & 0.05 & 16.36& 0.06& M70  \\
92.22 &      &     &  20.23& 0.21&  19.80 & 0.12 &      &     & M70  \\
97.27 &      &     &       &     &  $>$20.1&     &      &     & M70  \\
98.26 &      &     &       &     &  $>$20.0&    &      &      & M70  \\
99.24 &      &     &       &     &  $>$20.2&    &      &      & M70  \\
\hline
\end{tabular}
\end{table}

The maximum light in the $BVR$ bands was reached at about 
JD 2457072-73 (February 18-19), and perhaps some days earlier in the $I$ band.
The magnitudes at maximum are $B_{\rm max}=17.6$, $V_{\rm max}=16.65$,
$R_{\rm max}=16.15$, with about $\pm$0.1 mag uncertainty . 
The red colour of $(B-V)\approx1$ mag 
at maximum confirms the 
high value of interstellar reddening reported by Ochner et al. (2015).
For the first week past maximum the brightness declined quite slowly,
but then a very fast decline occurred. We can approximately estimate
the values of brightness decline for the first 15 days
after peak ($\Delta m_{15}$), which are usually used to describe
the rate of photometric evolution for SNe Ia and SNe of other
types with similar light curve shapes: $\Delta m_{15}(V)\approx2.6$,
$\Delta m_{15}(R)\approx2.5$. Such high values were observed only for three
among all well-studied SNe: SN 2002bj (Type Ib)(Poznanski et al., 2010),
2005ek (Ic)(Drout et al., 2013) and 2010X (Ic-pec)(Kasliwal et al., 2010).

Fig. 2 presents comparison of the $R$-band light curve of       
PSN J07285387+3349106 with the light curves for the three fastest
declining SNe listed above and also with  
the type Ibn SN 2010al (Pastorello et al., 2015), and one of the 
fastest decliners among SNe Ic SN 1994I (Richmond et al., 1996). 
The left panel compares the shape of the light curves which are
normalized to the peak. 
It is shown that the $R$-band
light curve of PSN J07285387+3349106 near maximum is matched by that for
SNe 1994I,
2002bj and 2010X, while SN 2005ek has sharper peak, and for SN 2010al
the peak is much broader. 
The fast decline after the peak is similar for SNe 2002bj, 2005ek, 2010X and
PSN J07285387+3349106.

The right panel shows the
absolute magnitude light curves.
We accepted the following values of distance moduli and total extinction 
$A_R$ for the SNe used for comparison: SN 1994I: 29.62, 1.05; SN 2002bj:
33.48, 0.2; SN 2005ek: 34.12, 0.44; SN 2010X: 34.0, 0.4; SN 2010al:
34.27, 0.15.   
For PSN J07285387+3349106 we
adopted a distance modulus of $\mu=33.84\pm0.15$ from the NED database,
which is calculated from the
redshift of the galaxy, corrected for Virgo and Great Attractor infall,
with $H_0=73$ km/sec/Mpc.
We adopted Galactic extinction $A_R=0.13$
(Schlafly \& Finkbeiner, 2011)
and the host galaxy interstellar reddening  $E(B-V)=1$ mag from 
Ochner et al. (2015). 
To calculate absolute magnitudes we need the value of 
$R_V=A_V/E(B-V)$. It is well known that for SNe $R_V$ may 
differ from the standard Galactic value of 3.1, and can be as
low as $\sim 1.5$, especially for SNe with high $E(B-V)$ 
(see e.g. Tsvetkov et al., 2014;
Wang et al., 2008). So we calculated absolute magnitudes using two
values of $R_V$: 3.1 and 1.5.
It is difficult to estimate uncertainty of the data on the host
galaxy extinction, as Ochner et al. (2015) does not provide such information.
Assuming this uncertainty to be about $10\%$, 
and adding in quadratures the errors of distance modulus ($\pm$0.15)
and peak magnitude ($\pm$0.1) 
we derive uncertainty
of absolute magnitude $\sim0.3$ mag. 
The right panel plot demonstrates that with $R_V=3.1$ PSN J07285387+3349106 
reaches peak magnitude of
$M_R=-20.15$ mag and is definitely
much brighter than all other SNe of our sample. 
If $R_V=1.5$,
than the maximum luminosity
of PSN J07285387+3349106 $M_R=-18.8$ mag is close to that for SNe 2010al
and 2002bj, it 
is higher than for SN 1994I by about 0.8 mag, and exceeds that  
for SNe 2005ek and 2010X by about 1.8 mag.

We conclude that PSN J07285387+3349106 belongs to a rare class 
of extremely fast declining SNe and is similar to 
SNe 2002bj, 2005ek and 2010X regarding the shape of the light curves.
The estimate of the peak luminosity 
for a heavily reddened SN depends strongly on the 
presently unknown value of $R_V$. We show that even if $R_V$ is
close to the minimal observed value, PSN J07285387+3349106 still
belongs to the brightest objects among SNe of similar classes. 

PSN J07285387+3349106 may represent an extreme case of fast declining
SNe, combining very fast decay and high maximum luminosity.  
Unfortunately, it is now too 
faint for our instruments, and we want to attract attention of
all interested observers to this object.   
We should note that for SNe 2002bj, 2005ek and
2010X there are practically no data on the radioactive tail of the 
light curve. PSN J07285387+3349106 presents the opportunity to 
obtain such data for the first time for a SN of this class.
Observations at largest telescopes are needed to help reveal the nature
of this object and for all class of fast declining SNe.  

\IBVSfig{10cm}{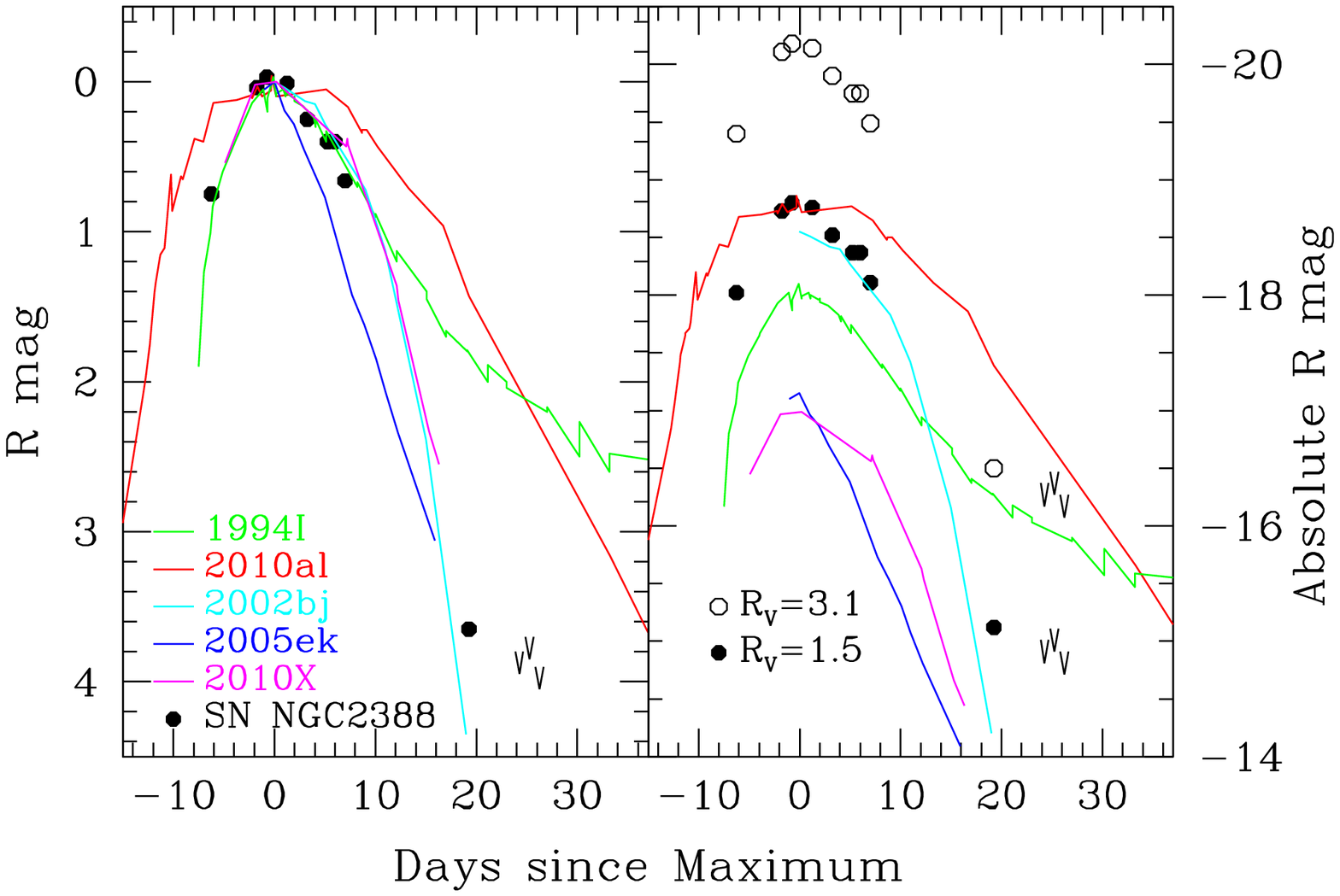}{Comparison of the $R$-band light curves for
 PSN J07285387+3349106 with those for 5 SNe.}
\IBVSfigKey{6140-f2.eps}{PSN J07285387+3349106}{light curve}

{\bf Acknowledgements:} This research has made use of the APASS database, 
located at the AAVSO web site. Funding 
for APASS has been provided by the Robert Martin Ayers Sciences Fund.
This research has made use of the NASA/IPAC 
Extragalactic Database (NED) which is operated by the 
Jet Propulsion Laboratory, California Institute of Technology, 
under contract with NASA. We are grateful to the anonymous referee
for important suggestions and comments.  

\references

Drout, M. R., Soderberg, A. M., Mazzali, P. A., et al., 2013, {\it ApJ}, 
{\bf 774}, 58

Henden, A. A., Levine, S. E., Terrell, D., Smith, T. C. \& Welch, D., 2012, 
{\it J. AAVSO}, {\bf 40}, 430

Kasliwal, M. M., Kulkarni, S. R., Gal-Yam, A., et al., 2010, {\it ApJ},
{\bf 723}, L98 

Ochner, P., Benetti, S., Pastorello, A., et al., 2015, {\it ATel}, {\bf 7105}  

Pastorello, A., Benetti, S., Brown, P. J., et al., 2015, {\it MNRAS}, {\bf 449}, 1921
 
Poznanski, D., Chornock, R., Nugent, P. E., et al., 2010, {\it Science}, {\bf 327}, 58

Richmond, M. W., VanDyk, S. D., Ho, W., et al., 1996, {\it AJ}, {\bf 111}, 327

Schlafly, E. F. \& Finkbeiner, D. P., 2011, {\it ApJ}, {\bf 737}, 103

Tsvetkov, D. Yu., Metlov, V. G., Shugarov, S. Yu.,  Tarasova, T. N., \&  
Pavlyuk, N. N., 2014, {\it CoSka}, {\bf 44}, 67
 
Wang, X., Li, W., Filippenko, A.V., et al., 2008,
{\it ApJ}, {\bf675}, 626

\endreferences

\end{document}